\documentclass[review,10pt]{elsarticle}
\usepackage{amssymb}
\usepackage{amsthm}
\usepackage{graphicx}
\usepackage{epsfig}
\usepackage{latexsym, amsmath}
\usepackage[english]{babel}
\usepackage{amsmath, amstext, amsfonts}
\usepackage{float}
\usepackage{caption}
\usepackage{subcaption}
\usepackage{appendix}

\def\R{{\mathbf R}}

\journal{}

\begin{document}
\pagenumbering{roman}
%\tableofcontents

\pagenumbering{arabic}
\begin{frontmatter}

\title{Multistable switching dynamical system with the location of symmetric equilibria}
%\title{Widening of the basins of attraction of a Bistable switching dynamical system with the location of symmetric equilibria}
\author[CARAO]{L.J.~Onta\~n\'on-Garc\'ia \corref{cor1}}
\ead{luis.ontanon@uaslp.mx}
\cortext[cor1]{Corresponding author}
\author[ipicyt]{E.~Campos-Cant\'on }
\ead{eric.campos@ipicyt.edu.mx}

\address[CARAO]{
  {\textsc{Coordinaci\'on Acad\'emica Regi\'on Altiplano Oeste},\\}
  {Universidad Aut\'{o}noma de San Luis Potos\'{\i}},\\
  \textsc{Kilometro 1 carretera a Santo Domingo, 78600,\\
  Salinas de Hidalgo, San Luis Potos\'{\i}, M\'exico
  \vskip 2ex}}

\address[ipicyt]{
 \textsc{Divisi\'on de Matem\'aticas Aplicadas,\\}
 {Instituto Potosino de Investigaci\'on Cient\'{\i}fica y Tecnol\'ogica A.C.}\\
 \textsc{Camino a la Presa San Jos\'e 2055 col. Lomas 4a Secci\'on, 78216,\\
 San Luis Potos\'{\i}, SLP, M\'{e}xico \vskip 2ex}}

\setcounter{page}{1}

\begin{abstract}

A switching dynamical system by means of piecewise linear systems in $\R^3$ that presents multistability is presented. The flow of the system displays multiple scroll attractors due to the unstable hyperbolic focus-saddle equilibria with stability index of type I, {\it{i.e.}}, a negative real eigenvalue and a pair of complex conjugated eigenvalues with positive real part. This class of systems is constructed by a discrete control mode changing the equilibrium point regarding the location of their states. The scrolls are generated when the stable and unstable eigenspaces of each adjacent equilibrium point generate the stretching and folding mechanisms to generate chaos, i.e., the unstable manifold in the first subsystem carry the trajectory towards the stable manifold of the  immediate adjacent subsystem.

The resulting attractors are located around four focus saddle equilibria. If the equilibria are located symmetrically to one of the axes and the distance between each equilibria is properly adjusted to generate two double-scroll chaotic attractors, the system can present from bistable to multistable solutions regarding the position of their initial states. In addition the resulting basin of attraction presents a significatively widening when the distance between the equilibria of the symmetric attractors is displaced.

\end{abstract}


\end{frontmatter}

\section{\label{sec:Introduction}Introduction}

Switching dynamical systems are characterized by the mutual existence of continuous and discrete dynamic behavior.
An interesting example of these systems is represented by piecewise linear (PWL) systems, particularly the ones described by dissipative systems with unstable dynamics which have been called \emph{unstable dissipative systems} (UDS) theory \cite{campos2,campos1,UDS4,capitulo_uds,campos_udsII}, which makes a characterization of the linear system onto two types depending on the spectrum of the linear operator. Considering switching systems with phase space in $\R^3$, the stability index of type I refers to one negative real eigenvalue and one pair of complex conjugated eigenvalues with positive real part.
If the location of the equilibria is adjusted taking into consideration this type of index, chaotic attractors with multi-scroll behavior can be generated.

This idea of multi-scroll attractor has been widely studied over the last decades. For example, in the work reported by Suykens in \cite{suykens93}, n-double scrolls in the Chua's system were presented. After that there have been different approaches to yield multi-scroll attractors. These approaches vary from modifying the nonlinear part in the Chua's system, to using nonsmooth nonlinear functions, such as hysteresis, saturation, threshold, step functions, fractional-order systems and chaos entanglement \cite{suykens93,sakthivel,sanchez,campos_TSO,campos_multiscroll,fengchen,jin}.
The main difference between the UDS of the type I technique of multi-scroll generation in contrast to some other techniques, is that for each equilibrium point introduced to the system a new scroll emerges, presenting systems with the same number of scroll that equilibrium points.

A well known property of chaotic systems is the highly sensibility  to initial conditions, that is, chaotic systems which are initialized with small differences in initial conditions will result in diverging trajectories. However, these trajectories are confined to the same attractor and in most cases the only one of the system. This fact has led the scientists to study and design systems that present more than one stable solutions resulting in two or more attractors given a fix set of parameters but different initial states. These multiple possible behaviors are isolated from each other and the term to refer to this property of the systems is defined as ``multistability" \cite{Arecchi,Sharma}. A common method to study this property is by means of the basins of attractions of the system, which basically  corresponds to the long-time response of the system due to the location of different initial conditions. Generally, this method  requires of exhaustive computation time depending on the system properties and both the  large scale of initial conditions and the time that the system is iterated \cite{Taborda,Wright}.

There have been several reported applications or natural occurrences about multistability. However, biology and electronics are two of the most recurrent areas. A few examples considering the former area are described in the multiple behavioral patterns in neural dynamics or the multistable coordination dynamics \cite{Briggman,Kelso}; in the dynamics of some biological central pattern generators through neurons connected in rings \cite{Canavier}; in the integrate-and-fire model of the neurons affected by white noise \cite{Foss}. Now, considering the case of the area of electronics, in the circuit implementation of the digital selection of an active set of neurons or in the use of nonlinear reactances or negative resistors \cite{Hahnloser,Kumagai}, switching dynamical systems describe interesting and common phenomena such as switching electrical circuits and systems involving both digital and analog components or physical systems affected by impact, sliding or friction forces \cite{goebel,haddad}, just to mention a few.

In this work, a new way to widen the basin of attraction of a bistable switching dynamical system considering only equilibria of the UDS type I which solution presents symmetric attractors will be described. This type of systems present an interesting relationship between the symmetric equilibria and the resulting basins of attraction related to each individual stable system, {\it i.e.}, if a specific size is considered the system will present bistable or multistable solutions.

The article is organized as follows: In Section \ref{sec:UDS}  the general theory is presented that envelops the generation of UDS; In Section \ref{sec:multiscroll}  the multiscroll attractors is introduced due to the variation of the distance between the commutation surfaces; Section 4 contains the generation of multistable systems with the widening of the basing on attraction. And finally conclusions are drawn in Section \ref{sec:Conclusions}.

\section{\label{sec:UDS}UDS theory}

Following the same structure as in \cite{UDS4,capitulo_uds,campos_udsII}, consider the class of switching linear system given by

\begin{equation}\label{ec_sistemA}
\dot{\mathbf{X}}=\mathbf{A}\mathbf{X} + \mathbf{B},
\end{equation}

\noindent
where $\mathbf{X}=[x_1,x_2,x_3]^T \in {\R}^3$ is the state vector,
$\mathbf{B}~=~[B_a,B_b,B_c]^T\in \R^3$ stands for a discrete real affine vector,
$\mathbf{A}~=~[a_{ij}]\in \R^{3\times 3}$ with $i,j=1,2,3$ denotes a nonsingular linear matrix.
The equilibria will be located at $X^*=-\mathbf{A}^{-1}\mathbf{B}$.
As described in Definition 2.1 in \cite{UDS4}, a system with stability index of the type I will be addressed as a system of the UDS type I. Besides, the following considerations have to be made in order to call Eq. (\ref{ec_sistemA}) an UDS  of type I that in addition generates an attractor $\mathfrak{A}$.
%%
%\begin{figure}[t!]
%  \centering
%%\includegraphics[width=5.cm]{UDSII}% Here is how to import EPS art
%        \includegraphics[width=0.32\textwidth]{Figuras/sistema1_xy}
%        \includegraphics[width=0.32\textwidth]{Figuras/sistema1_xz}
%        \includegraphics[width=0.32\textwidth]{Figuras/sistema1_yz}\\
%        \includegraphics[width=0.32\textwidth]{Figuras/sistema2_xy}
%        \includegraphics[width=0.32\textwidth]{Figuras/sistema2_xz}
%        \includegraphics[width=0.32\textwidth]{Figuras/sistema2_yz}
%\caption{\label{fig:UDSII} Projections of the attractor of the System A from eq. \eqref{ec_AB_R3} with \eqref{ec_rule2} onto the: a) $(x_1,x_2)$ plane; b) $(x_1,x_3)$ plane; c) $(x_2,x_3)$ plane, with initial condition $X_0=(1,0,0)$. Projections of the attractor of the System B from eq.\eqref{ec_AB_R32} with \eqref{ec_rule3} onto the: d) $(x_1,x_2)$ plane; e) $(x_1,x_3)$ plane; f) $(x_2,x_3)$ plane, with initial condition $X_0=(0,0,0)$. Results obtained from the numerical simulation of the systems.}
%\end{figure}

\begin{enumerate}
\item[a)]The linear part of the system must satisfy the dissipative condition $\sum_{i=1}^3\lambda_i<0$, where $\lambda _i, i=1,2,3$, are eigenvalues of $\mathbf{A}$. Consider also that one $\lambda_i$ is a negative real eigenvalue, and two $\lambda_i$ are complex conjugate eigenvalues with positive real part $Re\{\lambda_i\}>0$, resulting in a stable focus-saddle equilibrium $X^*$. This type of equilibria presents a stable manifold $M^s = span\{\lambda_1\}\in \R^3$ with a fast eigendirection and an unstable manifold $M^u= span\{\lambda_2,\lambda_3\}\in \R^3$ with a slow spiral eigendirection.

\item[b)]The affine vector $\mathbf{B}$ must be considered as a  discrete function that changes depending on which domain $\mathcal{D}_i\subset \R^3$ the trajectory is located. Accordingly  $\R^3=\cup_{i=1}^k\mathcal{D}_i$.
    Then a switching system based on the continuous linear system \eqref{ec_sistemA} and the discrete function $\mathbf{B}$ is given by:
\begin{equation}\label{ec_sistemABu}
\begin{array}{c}
\dot{\mathbf{X}}=\mathbf{A}\mathbf{\mathbf{X}} + \mathbf{B}(\mathbf{X}),\\\\
\mathbf{B}(\mathbf{X})=\left\{
          \begin{array}{ll}
            B_1, & \hbox{if $X\in \mathcal{D}_1$;} \\
            B_2, & \hbox{if $X\in \mathcal{D}_2$;} \\
            \vdots & \vdots     \\
            B_k, & \hbox{if $X\in \mathcal{D}_k$.}
          \end{array}
        \right.
\end{array}
\end{equation}

\end{enumerate}

The equilibria of system \eqref{ec_sistemABu} are $X^*_i=-\mathbf{A}^{-1}B_i$, with $i=1,\ldots, k$,  and each entry $B_i$ of the switching system  is considered in order to preserve bounded trajectories of system \eqref{ec_sistemABu}.

The commuting system given by Eq.~\eqref{ec_sistemABu} induces in phase space $\R^n$ the flow $(\varphi^t),{t\in \R}$ such
that each forward trajectory of the initial point $\mathbf{X}_0=\mathbf{X}(t=0)$ is the set $\{\mathbf{X}(t)=\varphi^t(\mathbf{X}_0):t\geq
0\}$. Furthermore, these systems have a dissipative bounded region ${\Omega}\subset \R^n$ named basin of attraction, such
that the flow $\varphi^t({\Omega})\subset {\Omega}$ for every $t\geq 0$. The attractor $\mathcal{A} $ is the largest attracting invariant subset of ${\Omega}$.

\section{\label{sec:multiscroll}Multiscroll switching systems}

Since the systems based on UDS type I present unstable equilibria, the design
of vectors $\mathbf{B}$ must be done considering the orientation of the stable and unstable
manifolds, $M^s$ and $M^u$, respectively, so that the trajectory of the system
remains contained in the attractor. The main idea is to match $M^u$ of the
first subsystem with the $M^s$ of the second one, as commented in \cite{capitulo_uds}. Now, in
order to generate a multiscroll switching systems, the following matrix $\mathbf{A}$
and vector $\mathbf{B}$ are given:

\begin{equation}\label{ec_AB_R3}
  \textbf{A}=\left(
    \begin{array}{ccc}
      0 & 1 & 0 \\
      0 & 0 & 1 \\
      -1.5&-1& -1 \\
    \end{array}
  \right),
\textbf{B}=\left(
    \begin{array}{c}
      0 \\
      0 \\
      B_c \\
    \end{array}
  \right).
\end{equation}
%\noindent

%By setting the parameters with $ \alpha_{31}=1.5,~\alpha_{32}=1,~\alpha_{33}=1$, the system
The spectrum of the matrix results in $\Lambda=\{-1.20, 0.10 \pm 1.11i\}$,  satisfying Definition 2.1 of \cite{UDS4} in which the system is an UDS of Type I. The component $B_c$  of the vector $B$ is governed by a discrete control mode (DCM), which can be designed  depending on the number of scrolls to be introduced. The DCM for which the system presents a  $2$ scrolls attractor commutes according to the value of $x_1$ as follows:

%The discrete affine vector $\mathbf{B}$ for which the system presents a  $2$ scrolls attractor commutes according to the value of $x_1$ as follows:

\begin{equation}\label{ec_rule2}
    B_c(x_1)=\left\{%
\begin{array}{ll}
    2, & \hbox{if $x_1\geq 1$;} \\
    1, & \hbox{otherwise.} \\
\end{array}%
\right.
\end{equation}

The equilibria of the system \eqref{ec_sistemABu} considering the matrix $\textbf{A}$ and vector $\textbf{B}$ defined in \eqref{ec_AB_R3} with \eqref{ec_rule2} are \mbox{$X^*_{p1}=(2/3,0,0)^T$} and $X^*_{p2}=(4/3,0,0)^T$, with the commutation surface located at the $x_1=1$ plane. Notice that the equilibria are only displacing along the $x_1$ due to they are given by $X^*=-\mathbf{A}^{-1}\mathbf{B}$. In case that a displacement along a different axis is required, different discrete affine vector $\mathbf{B}$ must be considered.

\begin{figure}[!t]
  \centering
\includegraphics[width=9cm]{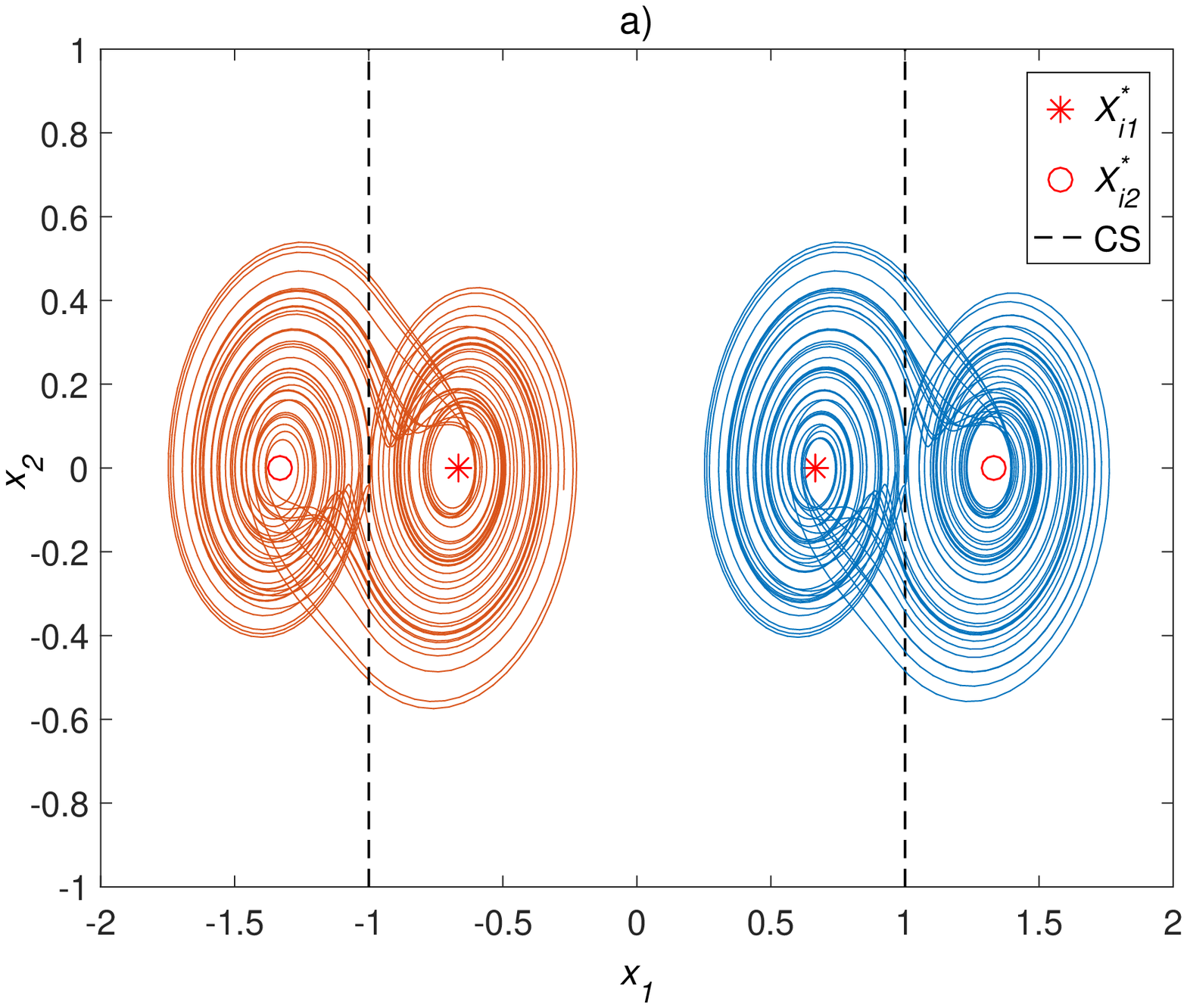}\\
\includegraphics[width=9cm]{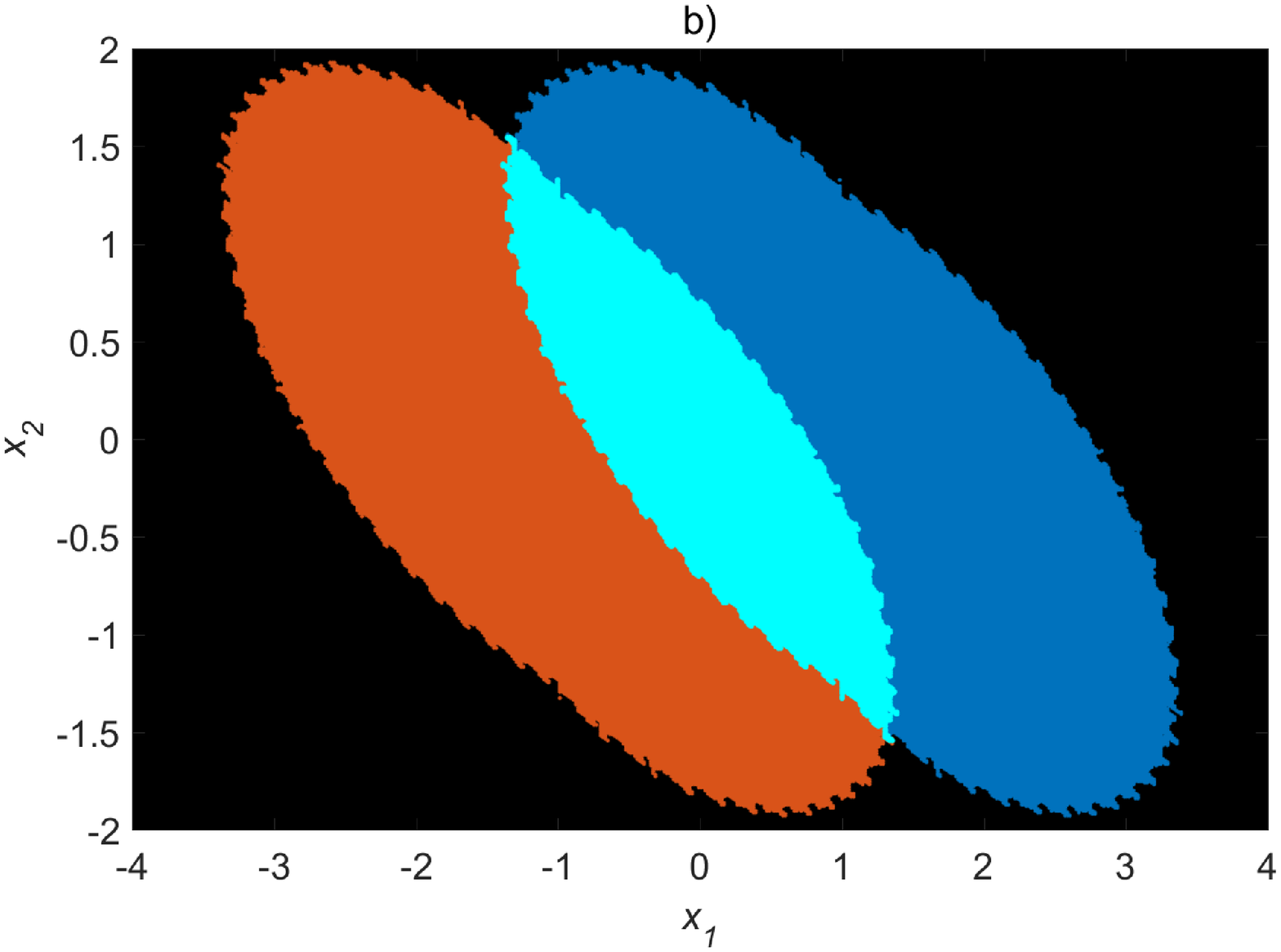}
%\caption{\label{fig:UDS_1_scroll} Projections of the attractors onto the $(x_1,x_2)$ plane generated by eq. \eqref{ec_AB_R3} and the
%SCL's: a) \eqref{ec_rule2}, and b) \eqref{ec_rulem2}. The commutation surfaces are marked with segmented line and the
%corresponding equilibria of each system is marked with an asterisk and and a circle.}
\caption{\label{fig:UDS_1_scroll} a) Projections of trajectories of the attractors onto the $(x_1,x_2)$ plane generated by eq. \eqref{ec_AB_R3} and the
DCM \eqref{ec_rule2} for the blue attractor in the right  side, and  \eqref{ec_rulem2} for the red attractor in the left side. The commutation surfaces are marked with segmented black lines and the corresponding equilibria of each system, $X^*_{i1}$ and $X^*_{i2}$ with $i=p,n$ marked with an asterisk and a circle for  \eqref{ec_AB_R3} with \eqref{ec_rule2} and \eqref{ec_rulem2}, respectively. b ) Basins of attraction generated by eq. \eqref{ec_AB_R3} and \eqref{ec_rule2} and \eqref{ec_rulem2} onto the $(x_1,x_2)$ plane. }

\end{figure}

By observing the righthanded side of Figure \ref{fig:UDS_1_scroll} a) which depicts the projection onto the $(x_1,x_2)$ plane of the attractor $\mathcal{A}_p$ generated by eq. \eqref{ec_AB_R3} with \eqref{ec_rule2}  marked with blue line, it can be appreciated that the system displays a double-scroll attractor centered at the $x_1=1$ plane marked with a segmented black line which stands as a commutation surface.  The equilibria of the
system $X_{p1}$ and $X_{p2}$ are depicted by a red asterisk and a circle, respectively. The initial condition from which the system was integrated by a fourth order Runge Kutta  method is $X_0=(1,0,0)^T$. The approximate size range of the attractor on each axis is given by:

\begin{equation}\label{ec_size}
  S_p = [R_1, R_2, R_3] ,
\end{equation}
\noindent where $R_i=\lvert \max(x_i)- min(x_i)\rvert$, with $i = 1,2,3$ results in $S_p\approx [ 1.5099,$ $1.0963,$ $1.0682 ]$  considering 4000 iterations of the integration method.
The largest Lyapunov exponent of the system was calculated by the approach described by Wolf {\it et. al.} \cite{wolf}, resulting in the positive value $\Lambda = 0.1765$ proving it is chaotic.

Now, if a different DCM is considered the system results in a completely different attractor. For example, if the discrete affine vector $\mathbf{B}$ commutes according to the value of $x_1$ as follows:

\begin{equation}\label{ec_rulem2}
    B_c(x_1)=\left\{%
\begin{array}{ll}
    -2, & \hbox{if $x_1\leq -1$;} \\
    -1, & \hbox{otherwise.} \\
\end{array}%
\right.
\end{equation}

Here the equilibria of the system \eqref{ec_sistemABu} considering the matrix $\textbf{A}$ and vector $\textbf{B}$ defined in \eqref{ec_AB_R3} with \eqref{ec_rulem2} are \mbox{$X^*_{n1}=-X^*_{p1}=(-2/3,0,0)^T$} and $X^*_{n2}=-X^*_{p2}=(-4/3,0,0)^T$, with the commutation surface located at the $x_1=-1$ plane. This results in a double-scroll attractor located in the negative side of $x_1$ as it can be appreciated in the lefthanded side of Figure  \ref{fig:UDS_1_scroll} a) with the projection onto the $(x_1,x_2)$ plane of a trajectory of the attractor $\mathcal{A}_n$ generated by eq. \eqref{ec_AB_R3} with \eqref{ec_rulem2}  marked with orange line; the initial condition is $X_0=(-1,0,0)^T$. Notice that the system is located in the negative region and it is symmetric to the positive one resulting in an attractor $\mathcal{A}_n$ with the same size range as $\mathcal{A}_p$, i.e, $S_n=S_p$. The largest Lyapunov exponent has the same value $\Lambda$ as it was  mentioned for the symmetric positive counterpart.
\begin{figure}[!t]
  \centering
\includegraphics[width=9cm]{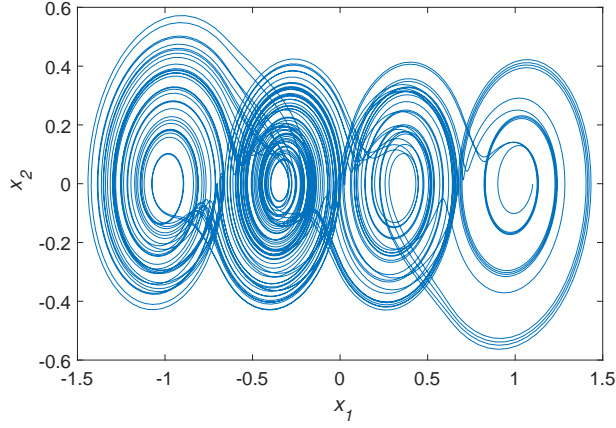}
\caption{\label{fig:UDS_4_scroll} Projection of a trajectory of the attractor generated by eq. \eqref{ec_AB_R3} with \eqref{ec_ruledelta} onto the $(x_1,x_2)$ plane with $\delta=1/2$.}
\end{figure}

Both systems present their respective basins of attractions $\Omega_p$ and $\Omega_n$ for  $\mathcal{A}_p$ and $\mathcal{A}_n$, respectively.
These basins can be appreciated in Figure \ref{fig:UDS_1_scroll} b), in which $\Omega_p$ is marked with blue and cyan colors, $\Omega_n$ is marked with orange and cyan colors. Every other point that is not contained on the basins is marked in black in a way that $\Omega_i^c=\{X_{m0}|\lim_{t\to \infty}\varphi^t(X_{m0})\not\subset \mathcal{A}_i\}$ and $\R^3=\Omega_i\cup \Omega_i^c$ with $i=p,n$. These basins are calculated considering as initial condition all the points in the plane $(x_1,x_2)$ with $x_3=0$ from $-4\leq  x_1 \leq 4$, $-2\leq  x_2 \leq 2$ spaced a distance of 0.04 and 0.08 from each point, respectively. The systems are iterated throughout 2000 iterations of the Runge Kutta of fourth order with a fix step of $1\times10^{-2}$.   Notice that both basins are symmetrical to each other and there is a small common region given by $\Omega_p\cap \Omega_n$, which is marked with cyan. These initial conditions $x_0\in\Omega_p\cap \Omega_n$ are well defined according to the DCM given by  \eqref{ec_rule2} or  \eqref{ec_rulem2}.
%The BA corresponds to the set of initial
%conditions whose long-time response (LTR) approaches the
%attractor

The Euclidean distances between equilibria of each system are $d_p(x^*_{p1},x^*_{p2})=d_n(x^*_{n1},x^*_{n2})=2/3$ corresponding to the attractors  $\mathcal{A}_p$ and $\mathcal{A}_n$, respectively. The commutation surface of each system is determined by a plane parallel to the plane $(x_2,x_3)$ and the middle point between equilibria. The Euclidean distances between the origin $O$ and the equilibria $x^*_{p1}$ and $x^*_{n1}$ are $d_{Op}(O,x^*_{p1})=d_{On}(O,x^*_{n1})=2/3$.
Now the idea is to generate a DCM based on \eqref{ec_rule2} and  \eqref{ec_rulem2} with the following characteristics:
\begin{itemize}
  \item The distances  $d_p(x^*_{p1},x^*_{p2})=d_n(x^*_{n1},x^*_{n2})=2/3$ remain constant.
  \item $0 < d_{Op}(O,x^*_{p1})=d_{On}(O,x^*_{n1})$.
  \item Three commutation surfaces: one of them at the origin and the others at  middle point between equilibria.
\end{itemize}
 The equilibria of the new system are located at  $X^*_i=-\mathbf{A}^{-1}B_i$, with $i=1,\ldots, 4$, notice that the location of the equilibria and the commutation surface between the scrolls are being displaced regarding the rates of $X^*=(\pm2/3B_c,0,0)^T$ and $x_1 = \pm2/3(B_c+1/2)$, respectively. Therefore, the system is defined by \eqref{ec_AB_R3} with the following DCM:

\begin{equation}\label{ec_ruledelta}
    B_c(x_1)=\left\{%
\begin{array}{ll}
    \delta+1,      & \hbox{if $ 2/3(\delta+1/2)\leq x_1 $;} \\
    \delta,       & \hbox{if $0\leq x_1 < 2/3(\delta+1/2)$;} \\
    -\delta,     & \hbox{if $ -2/3(\delta+1/2) \leq x_1 < 0 $;} \\
    -\delta-1,      & \hbox{if $x_1< -2/3(\delta+1/2)$;}
\end{array}%
\right.
\end{equation}

\noindent where $\delta\in \R^+$. This system  depending on the value assigns to $\delta$ can result in different forms, i.e., similar multiscroll attractors as depicted before, or  a four scroll attractor which will be addressed first.  The distances from the equilibria to the origin is given by $d_{Op}(O,x^*_{p1})=d_{On}(O,x^*_{n1})=2\delta/3$. In the particular case that $\delta=1/2$, the scrolls are generated around the equilibrium points $X^*_{p1,n1}=(\pm1/3,0,0)^T$ and $X^*_{p2,n2}=(\pm 1,0,0)^T$. The commutation surface  located at $x_1=0$ merges the previous attractors $\mathcal{A}_{p,n}$. This is depicted and better appreciated in Figure \ref{fig:UDS_4_scroll}. The Lyapunov exponent of the system is also equal to $\Lambda$, and the size range of the attractor $\mathcal{A}_{M}$ is $S_{M} \approx [2.8732,    1.1351,    1.1056]$ which is almost $90.29\%$ larger than $S_p$ and since the system presents only one stable solution, the basin of attraction $\Omega_{M}$ is unique and contains the attractor $\mathcal{A}_{M}$ generated with a system with four equilibrium points. This basin can be appreciated in Figure \ref{fig:UDS_1_cuenca} a) in which unlike the basins from the autonomous system in Figure \ref{fig:UDS_1_scroll} b), this one spans a wider basin on the plane $(x_1,x_2)$. To corroborate this fact and estimate the relative size of the basins of attraction \cite{Shrimali} we considered the following equation:

\begin{equation}\label{ec_basinsize}
    f=N_\mathcal{A}/N,
\end{equation}

\noindent where $N_\mathcal{A}$ corresponds to the number of initial conditions that trajectories asymptotically go to the attractor $\mathcal{A}$ regarding the full grid of points in $N$, both considered from the numerical simulation. The results of this can be appreciated in the Table \ref{Tab:sizes} where the values are evaluated through the integration method described above, considering a grid from $-10\leq x_1 \leq 10$ and $-10\leq x_2 \leq 10$ and $x_3=0$ with a spacing between initial condition of $0.04$, resulting in a grid of $500\times 500$ initial conditions. Notice in the first row for DCM \eqref{ec_rule2} that the size is  $f = 0.0252$ while for $\delta=1/2$ it is  $f = 0.2210$, which corresponds to an increase of $776.98\%$ !!!

This phenomenon can be better understood by taking into consideration the unstable and stable manifolds previously defined $M^u$ and $M^s$. When the double scroll attractor with eq. \eqref{ec_rule2} is initialized with any initial condition given slightly outside the righthanded side of the basin of attraction $\Omega_p$ considering $x_1>1$, the trajectory  with this initial condition oscillates leaving the domain due to $M^u$. Eventually this trajectory will cross the commutation surface at $x_1=1$ changing the dynamics and equilibrium point to the one on the lefthanded side. However the trajectory of the second subsystem initialized from this point in space, is located  far away and not in the proper direction of $M^s$ of the left side, now this trajectory cannot be attracted towards the center. Therefore, resulting in an increasing unstable oscillation that  cannon be bounded to the system's attractor. On the other hand, if the system had presented more equilibria at this side of the commutation law, the trajectory could have been  pulled due to the strength of each $M^s$ (regarding the position in which the trajectory is located) and bound the trajectory towards the location of the attractor, which in the system with eqs. \eqref{ec_ruledelta} for $\delta=1/2$ results in  the four scrolls attractor $\mathcal{A}_{M}$.

Thus, the relation of the system regarding the commutation law \eqref{ec_ruledelta} and its parameter $\delta$ prompts to the variation of the distance between the two double scroll attractor. This will be addressed next.

\section{Multistable switching system with symmetric equilibria}

\begin{figure}[!t]
  \centering
\includegraphics[width=0.45\textwidth]{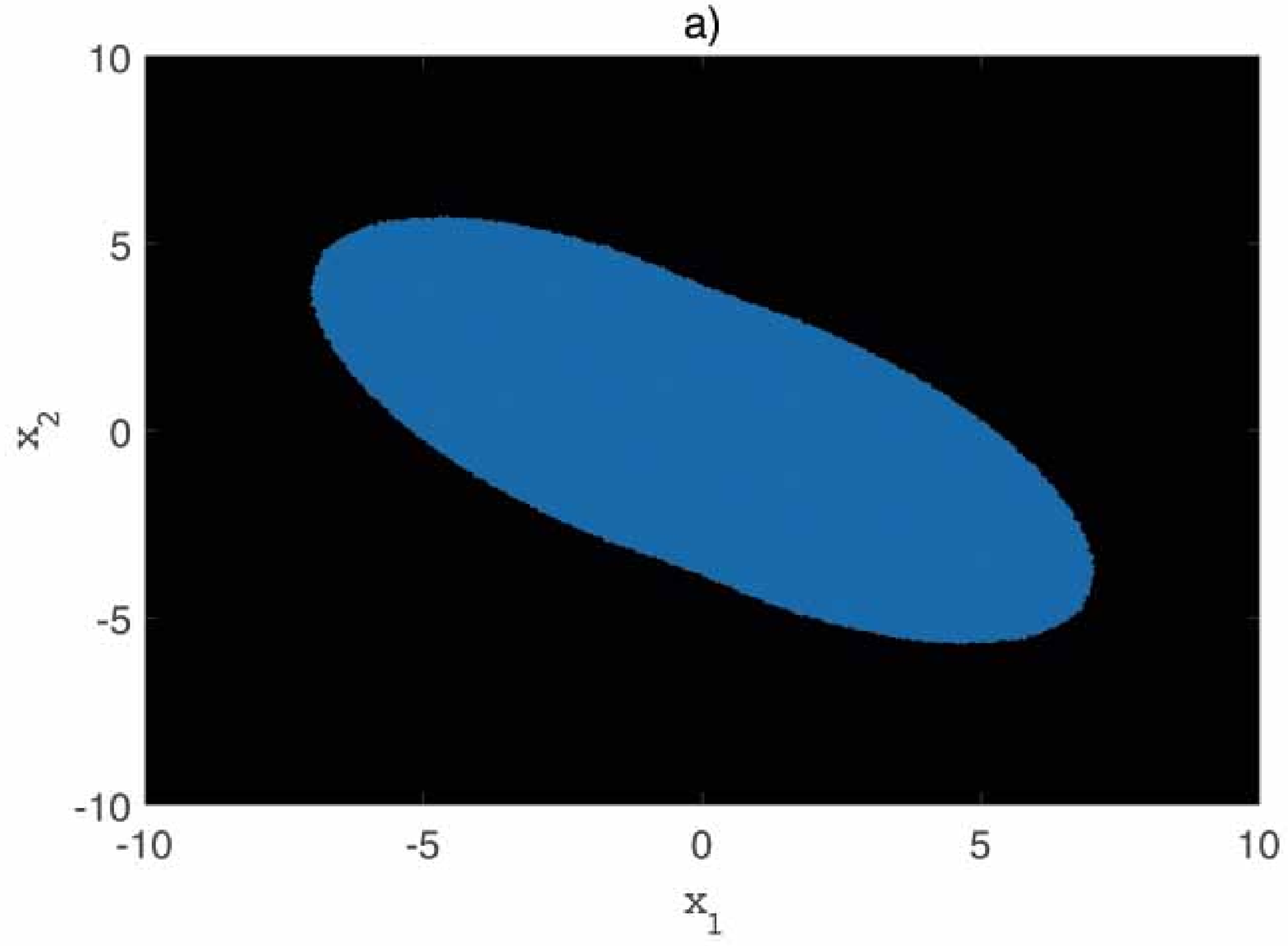}
\includegraphics[width=0.45\textwidth]{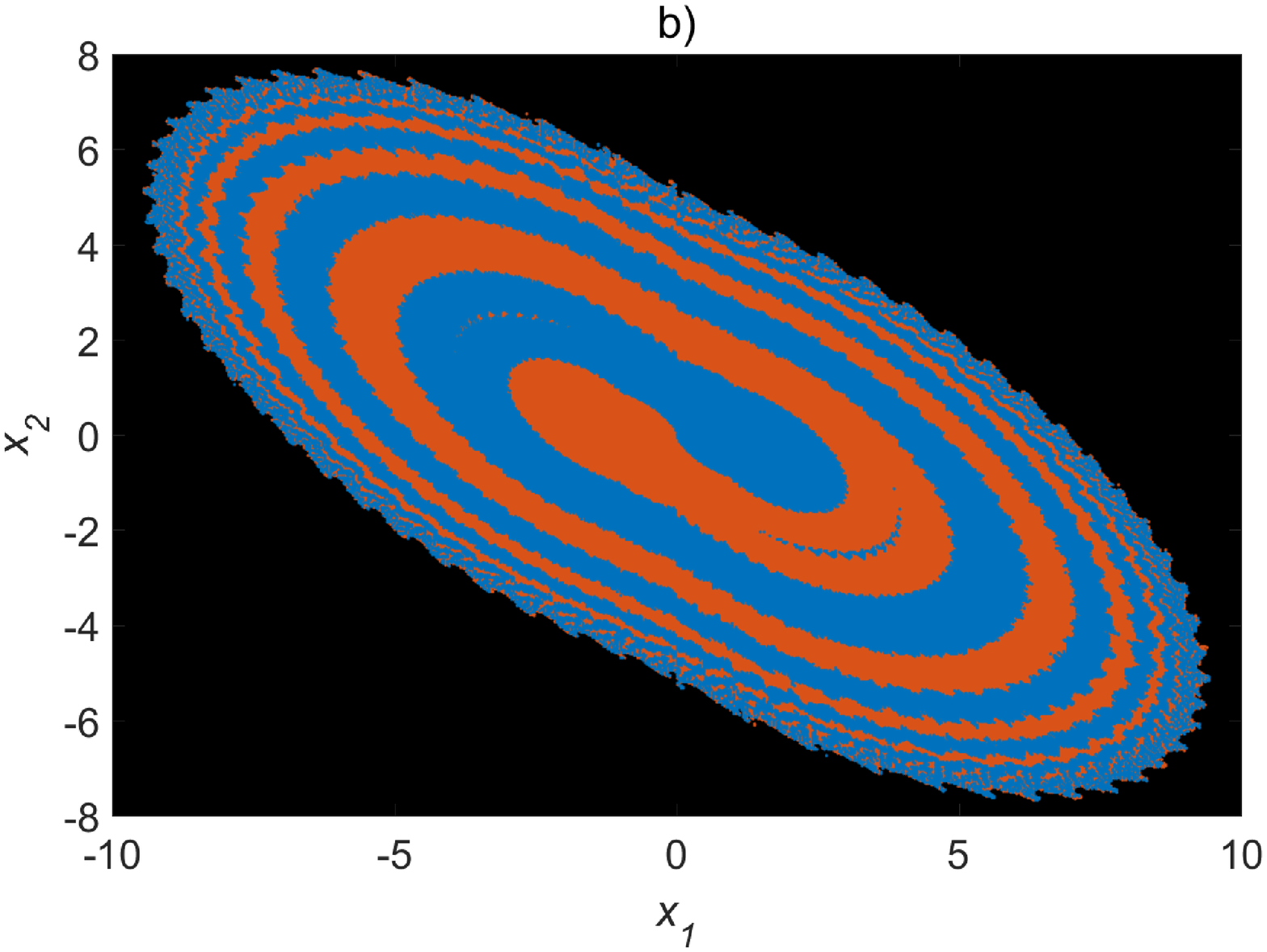}\\
\includegraphics[width=0.45\textwidth]{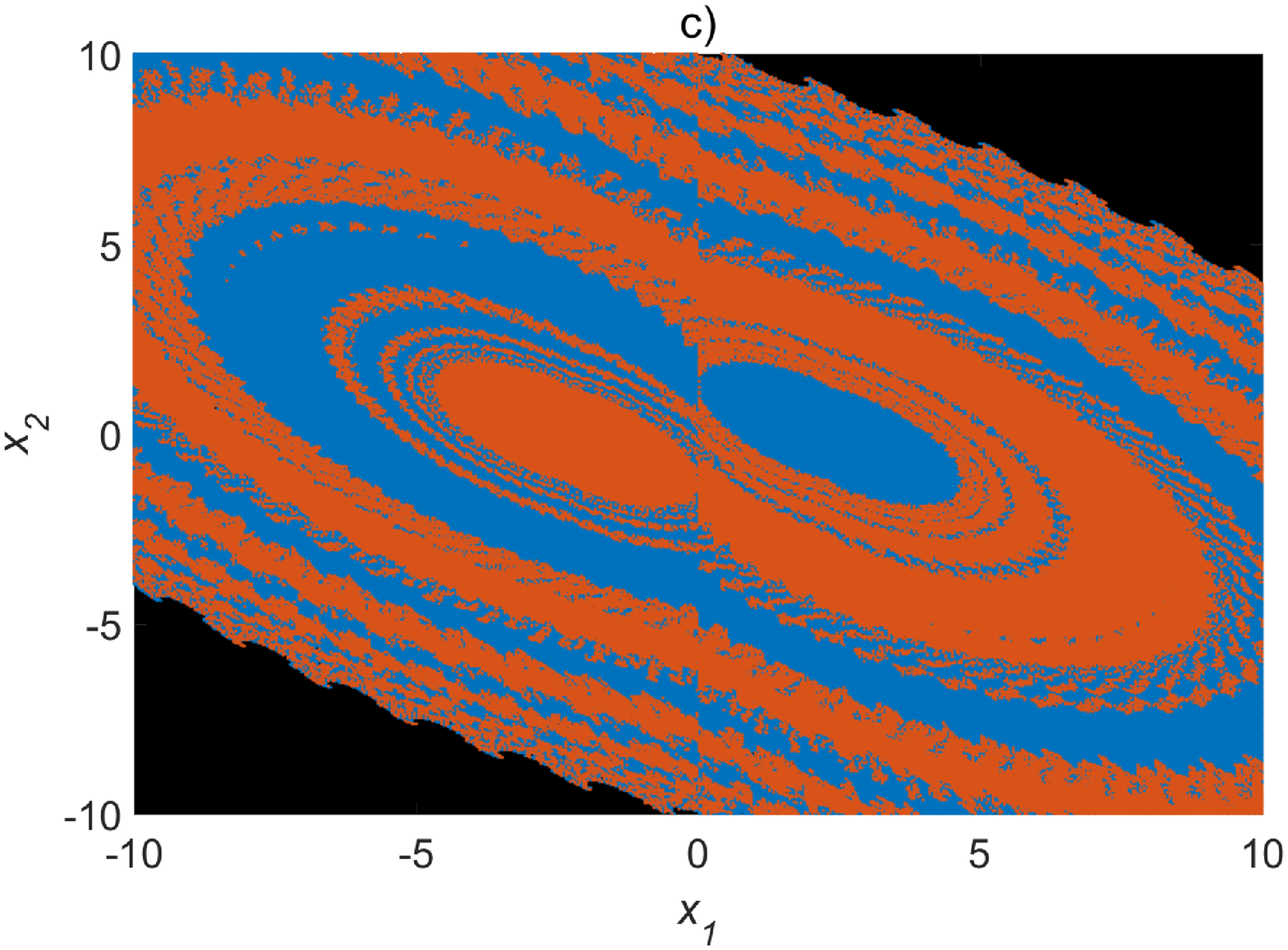}
\includegraphics[width=0.45\textwidth]{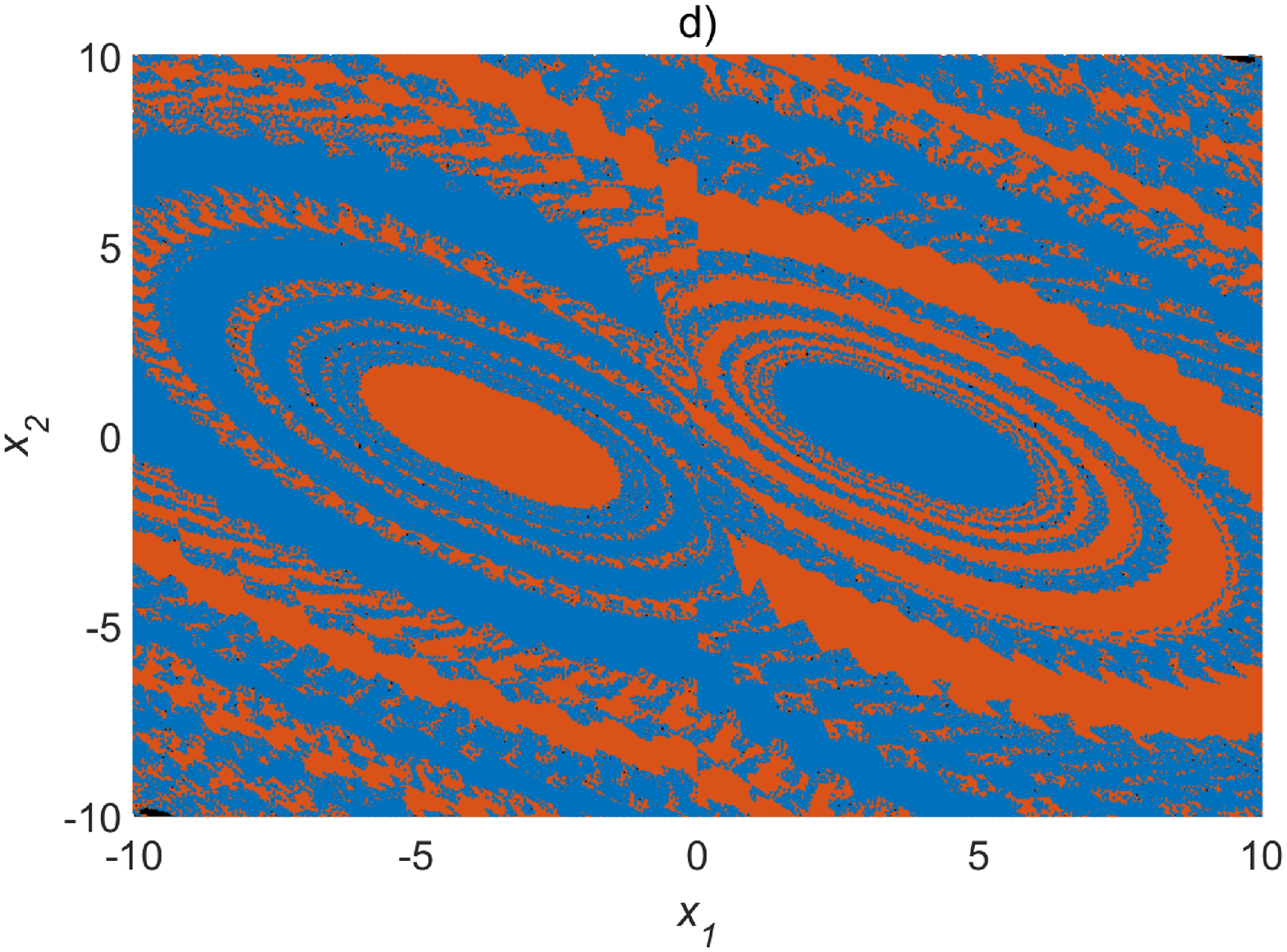}\\
\includegraphics[width=0.45\textwidth]{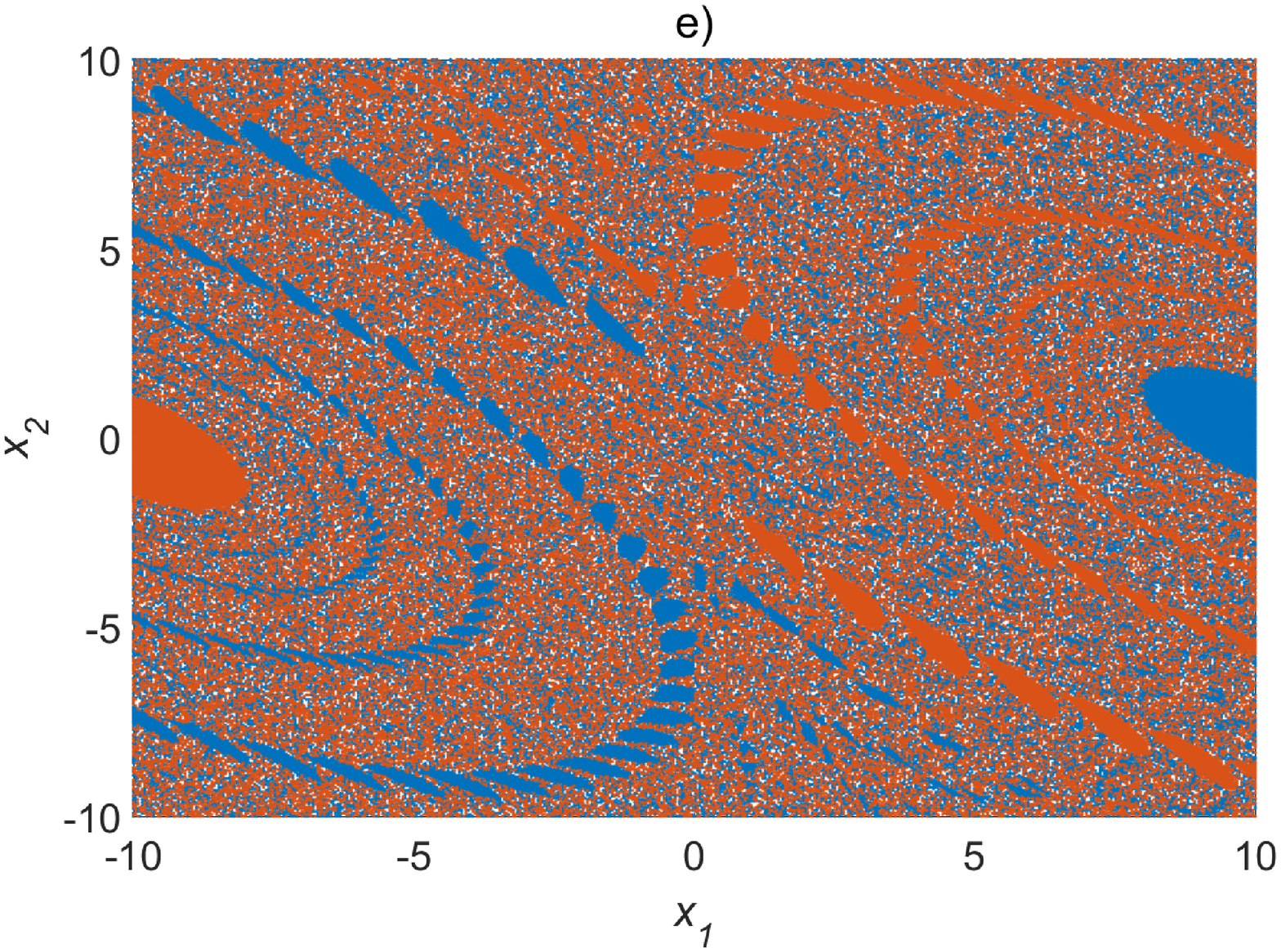}
\caption{\label{fig:UDS_1_cuenca} Section of the basins of attraction in the plane $x_3 = 0$ generated by eq. \eqref{ec_AB_R3} and \eqref{ec_ruledelta} with a) $\delta=1/2$; b) $\delta=1$; c) $\delta=3$; d) $\delta=5$ and e) $\delta=15$ onto the $(x_1,x_2)$ plane.}
\end{figure}

If a larger distance between a pair of equilibria of the two double-scroll attractors is considered, the attractors on the centered equilibria is able to oscillate freely and their natural size  fits in this space between the commutation surfaces. To exemplify this  the value of $\delta = 1$ is considered. With this discrete control mode the systems has four equilibrium points located at \mbox{$X^*_{p1,n1}=(\pm2/3,0,0)^T$} and $X^*_{p2,n2}=(\pm4/3,0,0)^T$, and  the three commutation surfaces are placed at  $x_1=-1,0,1$. Notice that this corresponds to the values of the two autonomous systems presented in the previous section with DCM's given by Eqs. \eqref{ec_rule2} and \eqref{ec_rulem2}. Now the multiscroll attractor has separated in a bistable mode resulting into two attractors, $\mathcal{A}_1$ and $\mathcal{A}_2$ (intending to avoid redundancy these attractors are not shown in the article  since they result in exactly the same as the ones depicted in Figure \ref{fig:UDS_1_scroll} a) for the separate systems). An interesting property of bistability is generated from the separation of the double scroll attractors in which each attractor is visited regarding the initial condition given, meaning that the basin of attraction is divided into subspaces which lead trajectories to each one of the attractors, or none of them whether the trajectory locates outside of any of the two basins of attraction escaping to infinity. This can be appreciated in Figure \ref{fig:UDS_1_cuenca} b) with basins of attraction $\Omega_1$ and $\Omega_2$ onto the plane $(x_1,x_2)$. Correspondingly $\Omega_1$ is marked in blue for $\mathcal{A}_1$, and $\Omega_2$ is marked in orange to $\mathcal{A}_2$. Notice that the spiral-like basins are non-convex and entangled among them  $ \Omega_1 \cup \Omega_2 \subset \R^3 $. It is also important to mention that $\Omega_1 \cap \Omega_2 = \emptyset$, meaning that there is no initial condition that results in both attractors.
Besides the complex structure and interaction between these basins of attraction, the property of intermingle and riddled basins that have been previously described in both discrete and continuous time systems as refereed in \cite{Shrimali,Camargo}, cannot be attribute in this case due to their formal definition.%Although the complexity of the basins and the fractal appearance in their borders (see Figures \ref{fig:UDS_1_cuenca} b)-e)), they are not riddled basins \cite{Alexander} due to the fact that the subspaces $\Omega_1$ and $\Omega_2$ are not smooth invariant near the commutation surfaces of the system.
The resulting size of each one of the attractors $\mathcal{A}_{1,2}$ and the Lyapunov exponent remain invariant to the size of $S_p$ and $\Lambda$, respectively.

In order to understand the dynamics of the bistable system represented by its basins of attractions let's consider some variations regarding the location of the equilibria and the resulting attractors due to $\delta$.

Three more values are considered,  $\delta = 3,5,15$. All three result in bistable attractors and part of their corresponding basins of attraction are depicted  onto the $(x_1,x_2)$ plane, see Figure \ref{fig:UDS_1_cuenca}  c), d) and e) respectively. Notice that when the distance increases the basins of attraction begin to expand in size and their structure begins to present more complex forms, i.e., the part considered of all basins of attraction (Figures b)-e)) present a small ball near the location of the corresponding attractor, and initial conditions far the attractors generate basins of attraction in complex spiral forms symmetrically, i.e., $\Omega_1\Leftrightarrow -\Omega_2$. All the resulting attractors have the same size $S_p$ and the same Lyapunov exponent $\Lambda$.

What doesn't remain the same is the size of each basin, which is increasing considerably with respect to the four scrolls attractor given considering $\delta=1/2$ in Figure \ref{fig:UDS_1_cuenca} a). To prove this, the relative size of the basins of attraction $f$ are also considered as they can be appreciated in Table \ref{Tab:sizes}. Here, the value of $f_{\delta_{1/2}}$ is greater than $f_{\delta_{1}}$. However, consider that this value only represents the attractor $\mathcal{A}_1$ on the positive size. The total size of the basins $\Omega_{1,2}$ is $2f_{\delta_{1}}$. Also notice that the value of $f_{\delta_{15}}$ is not apparently increasing as it is expected. This is because when the value of $\delta$ begins to increase, the trajectories of the system may take larger transitory states. With this value of $\delta$, the number of iterations for the systems instead of $2000$ is $6000$. However, not all initial conditions generate trajectories that go to the attractor due to unfinished transitory periods. This can be appreciated in the white spots marked on the Figure \ref{fig:UDS_1_cuenca} e) which are located among the basins.  A more exhaustive investigation of these points result in conditions that if the system is initialized with, the resulting trajectory will reach the attractor.

\begin{table}[]
\centering
\caption{Values of the relative sizes of the basins of attraction regarding $\delta$. All points  were considered from a grid of $-10\leq x_1 \leq 10$, $-10\leq x_2 \leq 10$ and $x_3=0$ with a spacing between initial condition of $0.04$, resulting in a grid of $500\times 500$ initial conditions. The values of $f$ only show the values of the basin which contains the attractor with $x_1>0$.}
\label{Tab:sizes}
\begin{tabular}{|c|c|c|c|}
\hline
\begin{tabular}[c]{@{}c@{}}Basin due\\ to $\delta$ \end{tabular} & \begin{tabular}[c]{@{}c@{}} $N_\mathcal{A}$ \end{tabular} & \begin{tabular}[c]{@{}c@{}}$N$\end{tabular} & $f$ \\ \hline
DCM \eqref{ec_rule2} & $6288$ & $500^2$ & $0.0252$ \\ \hline
$\delta_{1/2}$ & $55246$ & $500^2$ & $0.2210$ \\ \hline
$\delta_{1}$   & $49566$ & $500^2$ & $0.1983$ \\ \hline
$\delta_{3}$   & $107674$ & $500^2$ & $0.4307$ \\ \hline
$\delta_{5}$   & $122339$ & $500^2$ & $0.4894$ \\ \hline
$\delta_{15}$  & $94139$ & $500^2$ & $0.3766$ \\ \hline
\end{tabular}
\end{table}

\begin{figure}[!t]
  \centering
\includegraphics[width=9cm]{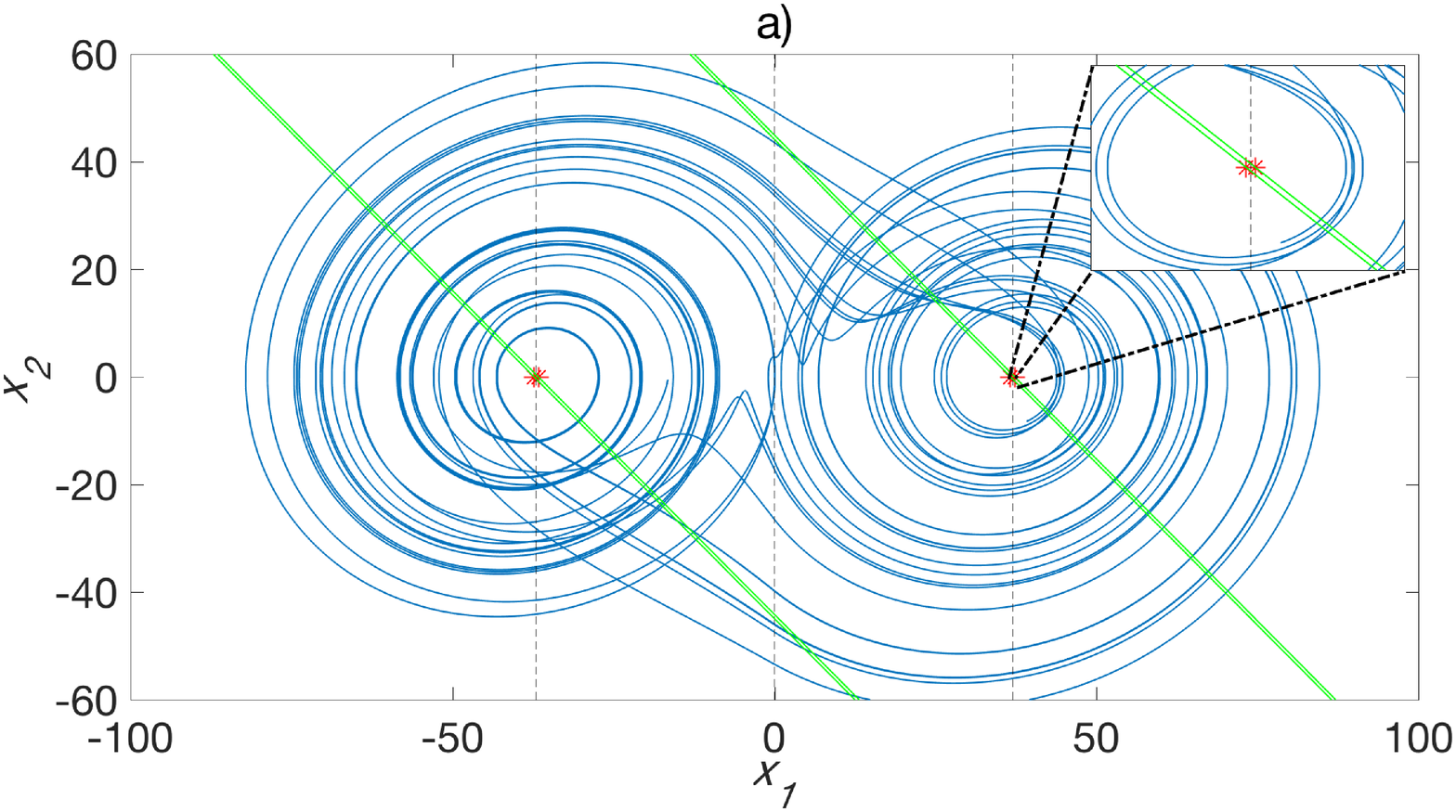}\\
\includegraphics[width=9cm]{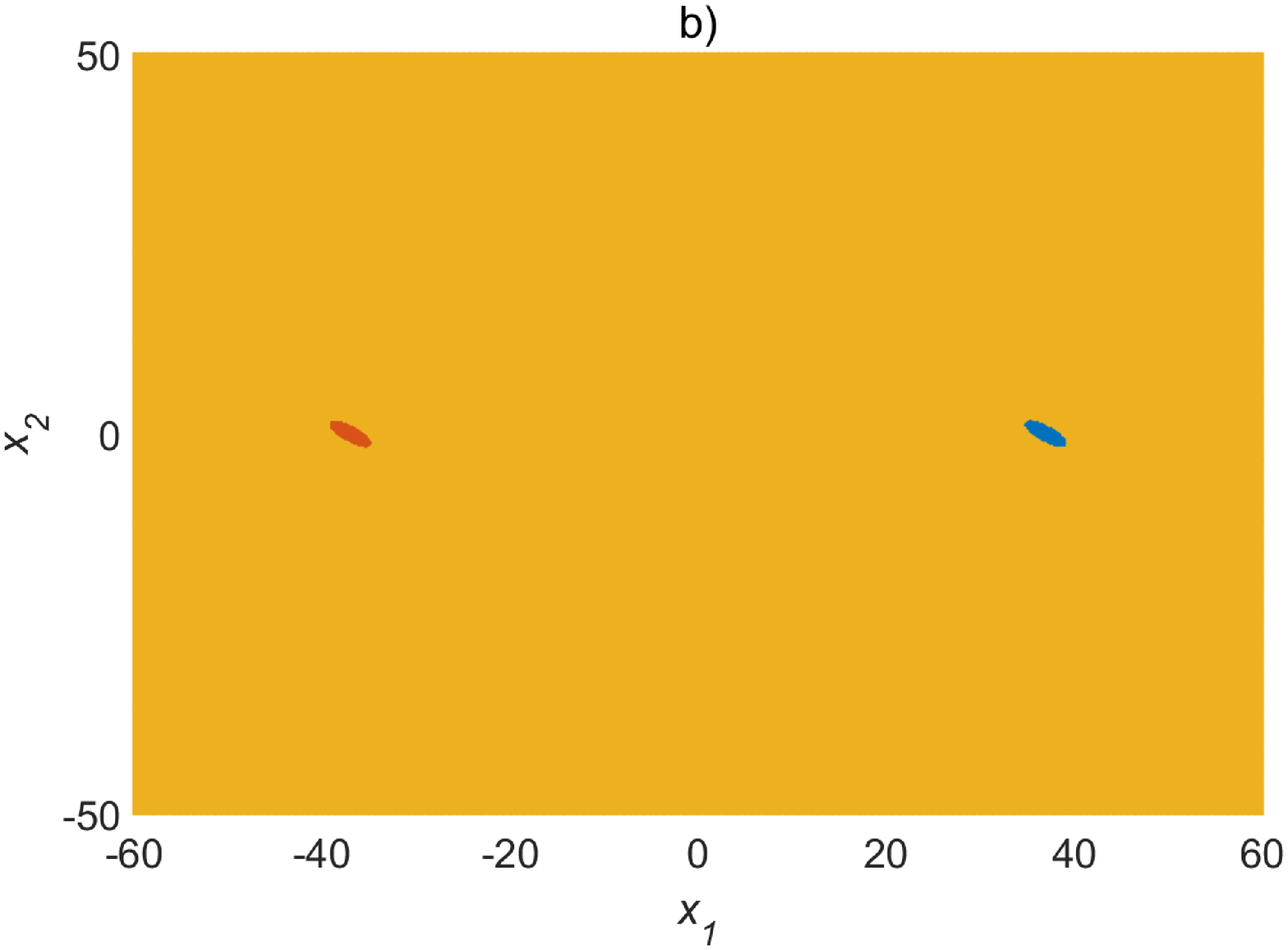}
\caption{\label{fig:UDS_1_cuencatri} a) Projection  onto the $(x_1,x_2)$ plane of the attractor of the system \eqref{ec_AB_R3} with \eqref{ec_ruledelta} and $\delta = 55$. The four equilibrium points of the systems are depicted in red asterisk, while the stable manifolds $M[s$ are depicted in green lines. The small box represents a zoom near the positive symmetric equilibria.
Section of the basins of attraction in the plane $x_3 = 0$ generated by eq. \eqref{ec_AB_R3} and \eqref{ec_ruledelta} with $\delta=55$ onto the $(x_1,x_2)$ plane. Marked in blue and red the initial conditions that asymptote the scroll near the symmetric equilibrium points, marked with yellow the conditions that asymptote the new larger scroll.}
\end{figure}

A natural question emerges at this point: what will happen to the bistable system if the parameter $\delta$ continues increasing?
The answer to this is rather easy to understand from the point of view of $M^s$ and $M^u$ as it is explained next.
Consider a larger value of $\delta$, for example $\delta=55$. With this value the symmetric equilibria located at $X^*_{p,n}=(\pm36.6666,0,0)^T$ are very far from each other comparing to the size of the attractor $S_p$, as it can be appreciated in Figure \ref{fig:UDS_1_cuencatri} a), where the projection onto the $(x_1,x_2)$ plane of a trajectory of the attractor of the system \eqref{ec_AB_R3} with \eqref{ec_ruledelta} and $\delta = 55$ is depicted. Notice the equilibria on the positive side marked with red asterisk, which due to the scale of the graphic these equilibria seems to be located almost in the same place. The stable manifolds (depicted in two extremely close green lines) corresponding to each equilibrium point are consider by the trajectory of the system as the continuation of the eigenvector, {\it i.e.}, when the trajectory is located between $0<x_1\leq37$ the stable manifold leading belongs to the the equilibria at the righthanded side  $X^*=(36.6666,0,0)^T$. When the system crosses the commutation surface at $x_1=37$, the trajectory is led by the stable manifold of the lefthanded side equilibria  $X^*=(37.6666,0,0)^T$.  This results in a  new larger basin of attraction encircling the bistable symmetric basins and resulting in a third stable state represented by a double-scroll attractor with larger dimension  (depicted in blue line) as it can be appreciated in Figure \ref{fig:UDS_1_cuencatri} a). The size of this new attractor results in $S_c\approx [ 223.3568, 164.3612, 160.3953]$.  However, the bistable attractors near the symmetric equilibria remain inside each symmetric basin of attraction located in the center region of the new larger stable solution. Thus a switching system with multistable solutions can be generated by increasing the distance between their symmetric equilibria $d_{Op}(O,x^*_{p1})=d_{On}(O,x^*_{n1})=2\delta/3$, which in addition presents also a larger basin of attraction that systems with small $\delta$. The basin of attraction of the system is presented in Figure \ref{fig:UDS_1_cuencatri} b). Notice the small basins of attraction inside the center balls corresponding to each scroll.

\section{\label{sec:Conclusions}Concluding remarks}
A method on how to generate multistable switching symmetric attractors from the UDS theory was presented. This systems can result in stable to multistable double scroll attractors when the distance between the equilibria of the symmetric subsystems is displaced. An important feature of this type of systems is that both the size of the basin of attraction and the size of the system increases considerably as the distance of the symmetric equilibria to the commutation surface in the origin augment.
This idea of augmented basin of attraction results as an interesting method to enlarge the possible initial condition given to a switching system in order that they fall into the attractor instead of become an unstable solution. Presenting an alternative to drive a bistable system to tristability.
Besides, applications in electronic systems can be improved using this method, since the possibilities of presenting an initial condition that asymptotes a stable state widen.

\section{Acknowledgements}
L.J.O.G. acknowledges the financial support through project PRODEP/DSA/103.5/15/6988 and UASLP for the financial support through  C15-FAI-04-80.80. E. Campos-Cant\'on acknowledges CONACYT for the financial support through project No. 181002.

%%%%%%%%%%%%%%%%%%%%%%%%%%%%%%%%%%%%%%%%%%%%%%%%%%%%%%%%%%%%%%%%%%%%%%%%%%%%%%%%%%%%%%%%%%%%%%%%%
%%%%%%%%%%%%%%%%%%%%%%%%%%%%%%%%%%%%%%%%%%%%%%%%%%%%%%%%%%%%%%%%%%%%%%%%%%%%%%%%%%%%%%%%%%%%%%%%%
%%%%%%%%%%%%%%%%%%%%%%%%%%%%%%%%%%%%%%%%%%%%%%%%%%%%%%%%%%%%%%%%%%%%%%%%%%%%%%%%%%%%%%%%%%%%%%%%%


\begin{thebibliography}{1}

\section*{References}

%1
\bibitem{campos2} E. Campos--Cant\'on, (2016). Chaotic attractors based on unstable dissipative systems via third-order differential equation. International Journal of Modern Physics C,  \textbf{27} (1), 1650008 (2016).

%2
\bibitem{campos1} E. Campos–-Cant\'on, I. Campos-–Cant\'on, J.S. Gonz\'alez--Salas and  F. Cruz--Ordaz, A parameterized family of single-–double-–triple-–scroll chaotic oscillations, Rev. Mex. Fis. \textbf{54}(6), (2008).
%3
\bibitem{UDS4}
    L. J. Onta\~n\'on--Garc\'ia, E. Jim\'enez-L\'opez, E. Campos--Cant\'on, M. Basin, A family of hyperchaotic multi-scroll attractors in $\R^n$, Appl. Math. Comput., \textbf{233}, (2014).

\bibitem{capitulo_uds}
L.J. Onta\~n\'on-Garc\'ia, E. Campos-Cant\'on, Bounded trajectories of unstable piecewise linear systems and its applications.
Advances in Mathematics Research, Nova Science Publishers, Inc., Vol. {\bf20}, April 01, pp. 149-172, ISBN: 978-163482742-3;978-163482741-6. (2015).


%5
\bibitem{campos_udsII}
    E. Campos-Cant\'on , R. Femat  \& Guanrong Chen. Attractors generated from switching unstable dissipative systems, CHAOS \textbf{22}, 033121, (2012)

%6
\bibitem{suykens93} J. A. K.  Suykens and J. Vandewalle, Generation of n-double scrolls $(n=1;2;3;4;...)$, IEEE Trans. Circuits Syst. I; {\bf40}(11), pp. 861--867, (1993).

%7
%\bibitem{cafagna} D. Cafagna and G. Grassi, New 3d-Scroll attractors in hyperchaotic Chua's circuits forming a ring,
% Int. J. Bifur. Chaos, {\bf13}(10),  pp. 2889–-2903, (2003).



%8
%\bibitem{yalsuyvan} M.E. Yal\c{c}in, J.A.K. Suykens and J. Vandewalle, Experimental confirmation of 3- and 5-scroll attractors from a generalized Chua's circuit, IEEE Trans. Circuits Syst., {\bf47}(3), pp. 425--429, (2000).



%9
\bibitem{sakthivel} G. Sakthivel, S. Rajasekar, K. Thamilmaran and S. K. Dana, Statistical measures and diffusion dynamics in a modified Chua's circuit equation with
multi-scroll attractors, Int. J. Bifur. Chaos, {\bf22}({1}), pp. 1--24 (2012).


\bibitem{sanchez}
C. S\'anchez-L\'opez, R. Trejo-Guerra, J. M. Mu\~noz-Pacheco and E. Tlelo-Cuautle,
N-scroll chaotic attractors from saturated function series employing CCII+s,
Nonlinear Dynamics, Volume {\bf61}, pp. 331--341, (2010).

%suykens93,sakthivel,sanchez,yalcin,campos_TSO,campos_multiscroll,fengchen

%12
%\bibitem{elwakil}
%        A. S. Elwakil, S. \"Ozoguz and M. P. Kennedy, A four-wing butterfly attractor from a fully autonomous system. Int. J. Bifur.
%    Chaos {\bf 13}(10), pp. 3093--3098, (2003).

%13
%\bibitem{yalcin}
%M.E. Yal\c{c}in, J.A.K. Suykens, J. Vandewalle and S. Ozoguz,
%Families of scroll grid attractors, Int. J. Bifur.
%Chaos {\bf12}(1), pp. 23--41, (2002).

%14
\bibitem{campos_TSO}
E. Campos-Cant\'on, I. Campos-Cant\'on, J. S. Gonz\'alez Salas and F.
Cruz Ordaz, A parameterized family of single-double-triple-scroll
chaotic oscillations, Rev. Mex. de F\'{\i}s., {\bf 54}, pp. 411--415, (2008).


%15
\bibitem{campos_multiscroll}E. Campos-Cant\'on, J.G. Barajas-Ram\'{\i}rez, G. Sol\'{\i}s-Perales and R. Femat, Multiscroll attractors by switching systems, Chaos {\bf20}, 013116, pp. 1–-6, (2010).



%16
\bibitem{fengchen}F. Chen, L. Xia, D. Guo and Y. Liu, A fractional-order multi-scroll chaotic system, Journal of Information \& Computational Science 10{\bf(4)},  pp. 1203–-1211, (2013).


\bibitem{jin} Zhang, H., Liu, X., Shen, X., \& Liu, J. (2013). Chaos entanglement: a new approach to generate chaos. International Journal of Bifurcation and Chaos, {\bf23}(05), 1330014.

\bibitem{Arecchi}
F.T. Arecchi, R. Badii \& A. Politi. {Generalized multistability and noise-induced jumps in a nonlinear dynamical system}. Physical Review A, {\bf32}(1), 402, 1985.


\bibitem{Sharma}
P.R. Sharma, M.D. Shrimali, A.  Prasad  \&  U. Feudel. {Controlling bistability by linear augmentation}. Physics Letters A, {\bf377}(37), 2329-2332, 2013.




%\bibitem{Pisarchik}
%A.N. Pisarchik  \& U. Feudel. {Control of multistability}. Physics Reports, {\bf540}(4), 167-218, 2014.




%\bibitem{Kapitaniak}
%B. Bla{\.z}ejczyk,  \& T. Kapitaniak. {Co-existing attractors of impact oscillator}. Chaos, Solitons \& Fractals, {\bf9}(8), 1439-1443, 1998.

\bibitem{Taborda}
J. A. Taborda, \& F. Angulo.,{Computing and Controlling Basins of Attraction in Multistability Scenarios.} Mathematical Problems in Engineering, Article ID 313154, (2015).

\bibitem{Wright}
J. A. Wright, J. H.Deane, M. Bartuccelli \& G. Gentile, {Basins of attraction in forced systems with time-varying dissipation}. Communications in Nonlinear Science and Numerical Simulation, {\bf29}(1), pp. 72-87. (2015).

\bibitem{Briggman}
Briggman, K. L., \& Kristan Jr, W. B. (2008). Multifunctional pattern-generating circuits. Annu. Rev. Neurosci., 31, 271-294.


\bibitem{Kelso}
Kelso, J. S. (2012). Multistability and metastability: understanding dynamic coordination in the brain. Philosophical Transactions of the Royal Society of London B: Biological Sciences, 367(1591), 906-918.


\bibitem{Canavier}
C. C. Canavier, D. A. Baxter, J. W. Clark \& J. H. Byrne, Control of multistability in ring circuits of oscillators. Biological cybernetics {\bf80}2, pp. 87-102. (1999)

\bibitem{Foss}
Foss, J., Moss, F., \& Milton, J. (1997). Noise, multistability, and delayed recurrent loops. Physical Review E, 55(4), 4536.


\bibitem{Hahnloser}
Hahnloser, R. H., Sarpeshkar, R., Mahowald, M. A., Douglas, R. J., \& Seung, H. S. (2000). Digital selection and analogue amplification coexist in a cortex-inspired silicon circuit. Nature, 405(6789), 947-951.





\bibitem{Kumagai}
Kumagai, S., \& Kawamoto, S. (1960). Multistable circuits using nonlinear reactances. Circuit Theory, IRE Transactions on, 7(4), 432-440.

\bibitem{goebel} R. Goebel, R.G.  Sanfelice, and A. Teel,  Hybrid dynamical systems, Control Systems, IEEE, {\bf29}(2), 28--93,(2009).

\bibitem{haddad} W.M. Haddad, V.  Chellaboina, S.G. Nersesov, Impulsive and hybrid dynamical systems, IEEE control systems magazine, (2006).
%

%
%\bibitem{orponen} P. Orponen, A survey of continuous-time computation theory, In Advances in algorithms, languages, and complexity (pp. 209--224). Springer US, (1997).
%
%%
%\bibitem{elwakil1}
%A.S. Elwakil, K.N. Salama and M.P. Kennedy, A
%system for chaos generation and its implementation in
%monolithic form, Proc. IEEE Int. Symp. Circuits and
%Systems (ISCAS 2000)(V), pp. 217--220, (2000).
%

\bibitem{UDSII}
Onta\~n\'on-Garc\'ia, L. J., Campos-Cant\'on, E., \& Femat, R. (2016). Analog Electronic Implementation of a Class of Hybrid Dissipative Dynamical System. International Journal of Bifurcation and Chaos, 26(01), 1650018.%

\bibitem{wolf}
A. Wolf, J.B. Swift, H.L. Swinney and J. Vastano, Determining Lyapunov exponents from a time series, Elsevier Science Publishers, Physica D, Vol. vol. {\bf16}, pp. 285--317, (1985).



\bibitem{Shrimali}
M. D. Shrimali, A. Prasad, R. Ramaswamy  \& U. Feudel,  The nature of attractor basins in multistable systems. International Journal of Bifurcation and Chaos, Vol. {\bf18}(06), pp. 1675-1688. (2008).

\bibitem{Camargo}
S. Camargo, R. L. Viana \& C. Anteneodo, Intermingled basins in coupled Lorenz systems. Physical Review E, Vol. {\bf85}(3), 036207. (2012).

\bibitem{Alexander}
J. C. Alexander, J. A. Yorke, Z. You \& I. Kan,  Riddled basins. International Journal of Bifurcation and Chaos, Vol. {\bf2}(04), pp. 795-813. (1992).


\end{thebibliography}
\end{document}